# Using Geometry to Rank Evenness Measures: Towards a Deeper Understanding of Divergence

Kawika Pierson[1]

[1] Willamette University, Salem Oregon; kpierson@willamette.edu

*Abstract*

While recent work has established divergence as a key framework for understanding evenness, there is currently no research exploring how the families of measures within the divergence-based framework relate to each other. This paper uses geometry to show that, holding order and richness constant, the families of divergence-based evenness measures nest. This property allows them to be ranked based on their reactivity to changes in relatively even assemblages or changes in relatively uneven ones.  We establish this ranking and explore how the distance-based measures relate to it for both order q=2 and q=1. We also derive a new family of distance-based measures that captures the angular distance between the vector of relative abundances and a perfectly even vector and is highly reactive to changes in even assemblages. Finally, we show that if we only require evenness to be a divergence, then any smooth, monotonically increasing function of diversity can be made into an evenness measure. A deeper understanding of how to measure evenness will require empirical or theoretical research that uncovers which kind of divergence best reflects the underlying concept.

INTRODUCTION

Alongside species richness, evenness is one of the most important and commonly used measures of biodiversity (Purvis and Hector 2000). Species evenness has been shown to be related to the productivity (Rohr et al. 2016; Zhang et al. 2012), healthy functioning (Maureaud et al. 2019), and stability of ecosystems (De Roy et al. 2013; Wittebolle et al., 2009). The rapidly developing streams of research interested in phylogenic and functional evenness (Le Bagousse-Pinguet 2021; Cadotte et al. 2010) are also important to recognize here, since measures of phylogenic and functional evenness are often based on measures of species evenness (Chao and Ricotta 2019; Ricotta et al. 2014; Mason et al. 2005). Despite this broadly recognized importance, or perhaps because of it, the topic of how best to measure evenness has been of perennial interest over more than 50 years and remains contentious to this day (Chao and Ricotta 2019; Kvålseth 2015; Tuomisto 2012; Jost 2012; Smith and Wilson 1996; Alatalo 1981; Pielou 1966).

Recently, Chao and Ricotta (2019) provided a unified framework that greatly advances our understanding of evenness. Central to this framework are the re-codification of the properties that all evenness measures must have, and a relaxation of linear value-validity (Kvålseth 2015) which enables evenness measures to be based on the divergence, rather than a strict distance, between vectors of relative abundances and the mean vector of perfect evenness. However, since their work presents the five divergence-based evenness measure families and the one distance-based family as equally valid measures of evenness, Chao and Ricotta (2019) provide no guidance on how to choose between their six families of measures, nor do they discuss whether these families relate to each other in important ways. This kind of guidance was quite influential for earlier conceptualizations of evenness (Tuomisto 2012; Smith and Wilson 1996) and providing similar guidance here is vital if ecologists want to avoid both the confusion that would result from a proliferation of measures, and the prospect of locking-in on a measure that later turns out to be non-optimal.

This paper shows that, holding order and richness constant, the families of divergence-based evenness measures nest within each other. This relationship allows us to rank them according to how reactive they are to changes in relatively even (more diverse) assemblages or changes in relatively uneven (less diverse) assemblages. While the seeds of this insight can certainly be found in the "Molinari shapes" of Smith and Wilson (1996), we extend the literature to illustrate this nesting and ranking geometrically in three-dimensions. Setting species richness, S, equal to 3 so that the geometry can be visualized, we graph each evenness measure as a surface, establishing that these measures only intersect at their minimum and maximum, and that their gradients are importantly differentiated. The appendix proves that these relationships hold for higher values of richness. We also illustrate the importance of this nesting by showing how each family of measures changes for simulated ecological data. That analysis reveals that the gap between the divergence-based measures increases with richness, making the case that a deeper understanding of which divergence best reflect evenness is a vital path for future research.

This geometric examination of evenness enables us to establish several previously unnoticed insights into the concept. First, we show that angular distance is a currently unexplored approach to measuring evenness. We use the framework of Chao and Ricotta (2019) to derive a family of angular distance-based evenness measures that are highly reactive to changes in relatively even assemblages. Second, we show that the divergence framework provides very little restriction on valid measures of evenness. We prove that, if Chao and Ricotta's (2019) divergence requirement is not strengthened, then any smooth, monotonically increasing function of diversity can be made into a measure of evenness by following their approach. Finally, we build on these insights to provide suggestions for how we might better identify or define the concept of evenness in future work.

## THE CURRENT UNDERSTANDING OF EVENNESS

Diversity and evenness are typically measured using the relative abundances of the species in an assemblage, known as the vector of species relative abundances (Ricotta 2003). Over time dozens of evenness measures have been proposed (Smith and Wilson 1996), yet aside from the original work in information theory that unifies all measures of diversity (Hill 1973), there was, until recently, no unified framework we could use to understand evenness measures in general. Chao and Ricotta (2019) move the field of evenness measurement forward substantially by providing such a framework. They modify the approach of Kvålseth (2015) to define the requirements of an evenness measure as:

>1a: The principle of transfer – Transferring abundance from a dominant species to a rare species should increase evenness.
>
>1b: Continuity – Evenness should be continuous and symmetric (invariant to permutations of the relative abundances).
>
>2: Rarity – Adding a rare species without changing diversity should not increase evenness.
>
>3a: Independence – Evenness should take a fixed interval regardless of richness and diversity.
>
>3b: Scale Invariance – Evenness should not depend on the units used, so that the evenness of raw abundances will be the same as the evenness of relative abundances.

The central difference between these requirements and the requirements of Kvålseth (2015) is his concept of value-validity, which argued that an evenness measure should capture the Euclidean distance between the vector of relative abundances and a completely even assemblage, called the mean vector. Rather than adopt this requirement, Chao and Ricotta (2019) instead relax it by defining evenness measures as divergences from the mean vector rather than strict distances. This is an important innovation in their work and enables all evenness measures to be combined within a unified framework.

This approach results in five classes of divergence-based evenness measures, defined as functions of their order, q, diversity (Hill 1973) of order q, $^qD$, and the species richness of each assemblage, S:

$$\text{Class 1: } {}^qE_1 = \frac{1 - ({}^qD)^{1-q}}{1 - S^{1-q}}$$

$$\text{Class 2: } {}^qE_2 = \frac{1 - ({}^qD)^{q-1}}{1 - S^{q-1}}$$

$$\text{Class 3: } {}^qE_3 = \frac{{}^qD - 1}{S - 1}$$

$$\text{Class 4: } {}^qE_4 = \frac{1 - (1/{}^qD)}{1 - (1/S)}$$

$$\text{Class 5: } {}^qE_5 = \frac{\ln({}^qD)}{\ln S}$$

There is also one distance-based evenness measure, which is written as a function of the proportions p, richness S, and an order q (originally α, but there is little practical distinction).

$$\text{Class 6: } {}^qE_6 = 1 - \left[\frac{\sum_{i=1}^{S}|p_i - (1/S)|^q}{[1 - (1/S)]^q + (S-1)S^{-q}}\right]^{1/q}$$

Chao and Ricotta (2019) restrict their analysis to measures of order 1 (where these expressions reduce to special forms) and measures of order 2, arguing that higher order measures focus too much on the abundances of dominant species. Beyond this though, they give no direction on how one should choose between these measures, and do not discuss whether or how these measures relate to each other.

## A GEOMETRIC VIEW OF EVENNESS

*Divergence Based Measures of Evenness when q=2*

Discussing evenness using analytic expressions is expositionally compact, but illustratively opaque. It is much easier to understand the relationships between these measures if we instead view them geometrically. To begin, the vector of relative abundances is a vector in S dimensional space. For visualization purposes we will set S equal to 3 so that we can explore these relationships in three dimensions, but the appendix shows that our results hold for higher S.

Figure 1

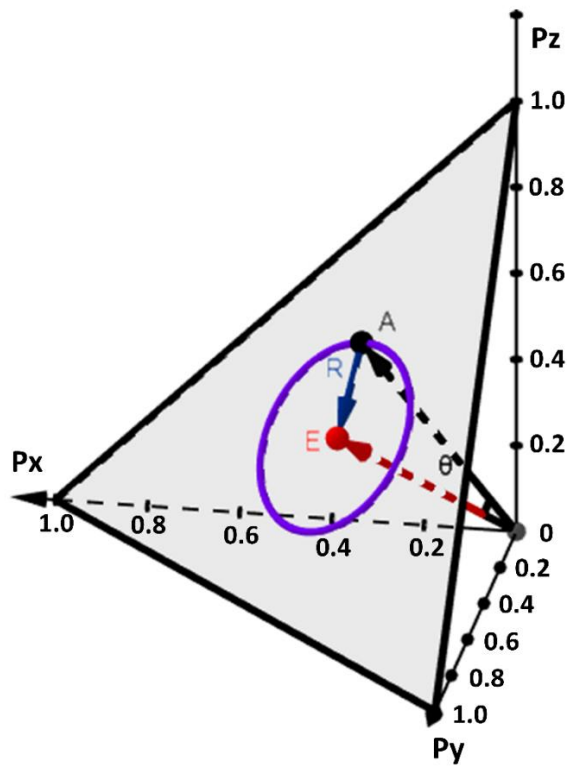

Figure 1. The triangle of valid relative abundances is shaded in this figure, with the mean vector **E** and a vector of valid relative abundances **A** forming a right triangle with their distance **R** = **A**-**E**. The angle between the two vectors is θ. A set of points including **A** and equidistant from **E** is drawn as a circle. Abundances on any circle equidistant from **E** have equal diversity and equal evenness when q=2 providing this order with perfect rotational symmetry.

There is a constraint on valid relative abundance vectors because their components must sum to 1. All valid vectors of relative abundance end on an S-1 dimensional regular simplex, which in the case where S=3 is the equilateral triangle formed by the plane x+y+z = 1 where x<1, y<1, and z<1. When order q=2 every point on this triangle that is an equal distance from the end of the mean vector **E** will have equal diversity. By extension, every equidistant point will also have equal evenness (measured using a single class). Thus, for q=2 both evenness and diversity display perfect rotational symmetry around the mean vector. Figure 1 shows this geometry.

Focusing on order 2 simplifies things because when q=2 class 1 is equivalent to class 4, and class 2 is equivalent to class 3. Thus, the divergence-based measures when q=2 are:

$$\text{Class 1,4: } {}^2E_{1,4} = \frac{1 - (1/{}^2D)}{1 - (1/S)}$$

$$\text{Class 2,3: } {}^2E_{2,3} = \frac{{}^2D - 1}{S - 1}$$

$$\text{Class 5: } {}^2E_5 = \frac{\ln({}^2D)}{\ln S}$$

Every divergence-based evenness measure is expressed as a function of diversity. The diversity of order 2 of a vector is the multiplicative inverse of the square of its magnitude. Or:

$$ {}^2D = \frac{1}{\|\vec{A}\|^2} = \frac{1}{\sum_{i=1}^{S}(p_i)^2}$$

$$\text{if } \mathbf{A} = (p_x, p_y, p_z) \text{ then } {}^2D = \frac{1}{p_x^2 + p_y^2 + p_z^2}$$

Further, since the sum of the components of **A** must be 1, diversity depends on just $p_x$ and $p_y$:

$$^2D = \frac{1}{2p_x^2 + 2p_y^2 + 2p_xp_y + 1 - 2p_x - 2p_y}$$

Each divergence based evenness measure is a transformation of this diversity surface. Class 2,3 evenness ($^2E_{2,3}$) simply shifts diversity down and scales it to fit between 0 and 1:

$$^2E_{2,3} = \frac{1}{2} \cdot \left( \frac{1}{2p_x^2 + 2p_y^2 + 2p_xp_y + 1 - 2p_x - 2p_y} - 1 \right)$$

class 1,4 evenness ($^2E_{1,4}$) is the scaled multiplicative and additive inverse of diversity:

$$^2E_{1,4} = \frac{3}{2} \cdot \left[ 1 - \left( 2p_x^2 + 2p_y^2 + 2p_xp_y + 1 - 2p_x - 2p_y \right) \right]$$

and class 5 evenness ($^2E_5$), is a logarithmic transformation of diversity:

$$^2E_5 = \frac{1}{\ln(3)} \cdot \ln\left( \frac{1}{2p_x^2 + 2p_y^2 + 2p_xp_y + 1 - 2p_x - 2p_y} \right)$$

Figure 2 plots each of these surfaces in the same space, along with a two dimensional view enabled by the rotational symmetry of order q=2 evenness. This gives us a novel insight into how these measures relate to each other by showing that, holding order and richness constant, the divergence-based evenness measures nest within each other. At their minimum and maximum these measures are equal, but otherwise the three surfaces never intersect.

This nesting occurs because the rates of change of these measures as a function of their distance from the mean vector are importantly different. For changes in evenness that are very close to the mean vector, and therefore occur within relatively even assemblages, $^2E_{1,4}$ always diverges the least, $^2E_{2,3}$ always diverges the most, and $^2E_5$ always diverges by an amount somewhere between these two. Close to the vertices of the triangle of valid abundances, where the changes reflect shifts in uneven assemblages, these relationships flip, with $^2E_{2,3}$ diverging the least, $^2E_{1,4}$ diverging the most, and $^2E_5$ still diverging by a moderate amount. Thus, $^2E_{1,4}$ is most reactive to changes in relatively uneven assemblages, $^2E_{2,3}$ is most reactive to changes in relatively even assemblages, and $^2E_5$ is somewhere between these two.

Figure 2

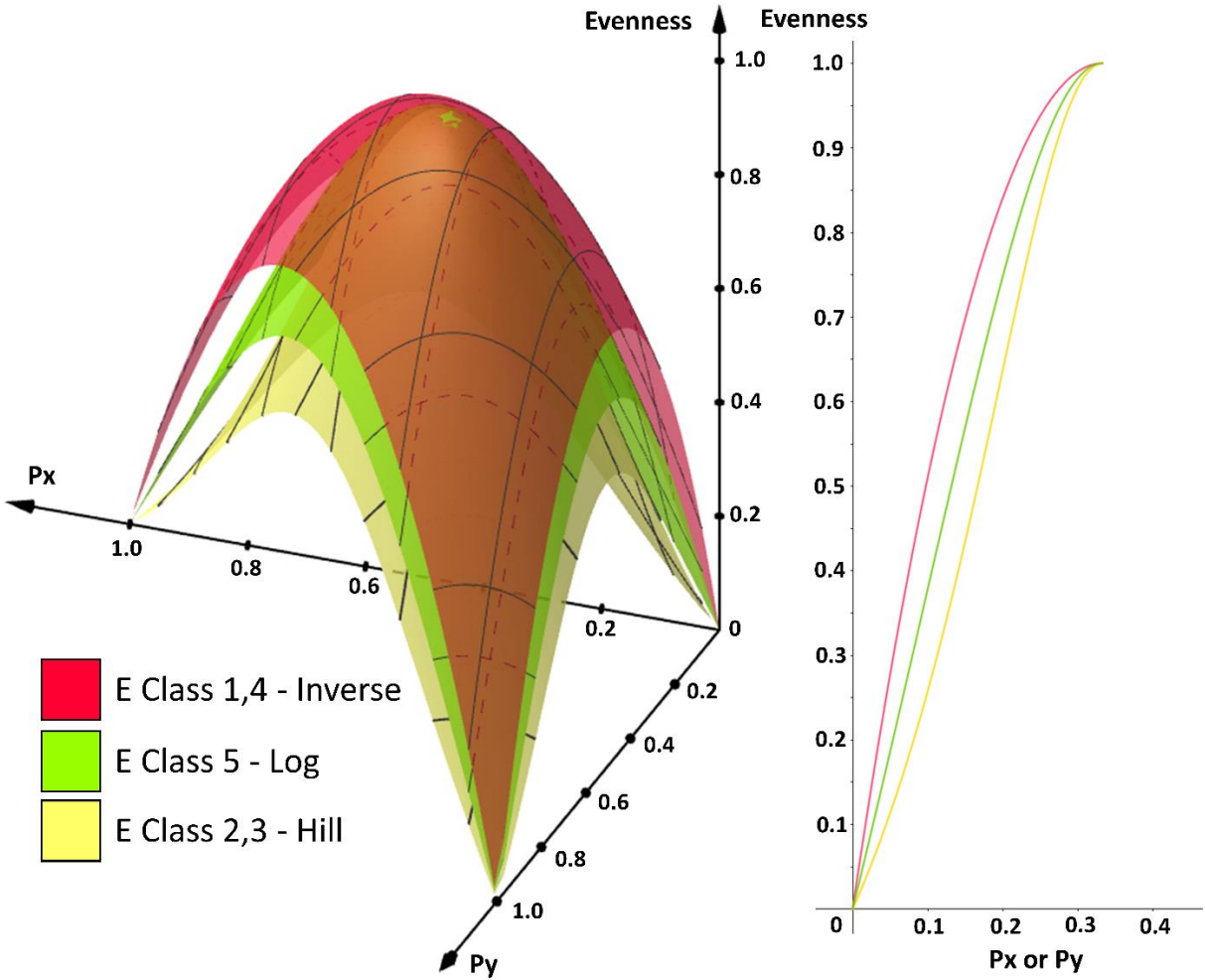

Figure 2. To the left, each of the divergence-based evenness measures of order q=2 is plotted as a surface in three-dimensional space. These measures nest within each other, only intersecting at their maximum and minimum. To the right each of the divergence-based evenness measures of order 2 is represented by a line, showing their values for the case where the x axis is equal to both the x and y components of the relative abundance vector.

This can be expressed as a ranking with respect to each measure's reactivity to changes in relatively uneven assemblages:

$$^2E_{1,4} > \,^2E_5 > \,^2E_{2,3}$$

with this ranking being reversed for reactivity to changes in relatively even assemblages.

Of course, the full relationship between these surfaces is determined by their partial derivatives with respect to diversity, but because the surfaces only intersect at their minimum and maximum, the fact that $^2E_{1,4}$ is always above $^2E_{2,3}$ and $^2E_5$ is sufficient to prove that $^2E_{1,4}$ diverges less for changes in high evenness assemblages and then "catches up" by diverging more for changes in low evenness assemblages.

*Distance-Based Measures of Evenness when q=2*

The Euclidean distance measure of Kvålseth (2015), which we call $^2E_6$, can also be transformed into a surface:

$$^2E_6 = 1 - \frac{\sqrt{\left(p_x - \frac{1}{3}\right)^2 + \left(p_y - \frac{1}{3}\right)^2 + \left(1 - p_x - p_y - \frac{1}{3}\right)^2}}{\sqrt{\frac{2}{3}}}$$

This surface is a cone because distance diverges linearly as one moves away from the mean vector.

The Euclidean distance is not the only valid way to think about the distance between the vectors **A** and **E**, however. As Figure 1 shows, these vectors form a right triangle, and so another candidate for the distance between the hypotenuse **A** and the adjacent side **E** is the angle θ.

The appendix follows the framework provided by Chao and Ricotta (2019) to derive a class of evenness measures using the concept of angular distance, which we call $^qE_7$:

$$^qE_7 = 1 - \frac{\cos^{-1}\left[\frac{(^qD)^{1/q}}{(S)^{1/q}}\right]}{\cos^{-1}\left[\frac{1}{(S)^{1/q}}\right]}$$

$^qE_7$ can also be written as a surface when q=2 and S=3:

$$^2E_7 = 1 - \frac{\cos^{-1}\left[\dfrac{1}{\sqrt{3}\cdot\sqrt{2p_x^2 + 2p_y^2 + 2p_xp_y + 1 - 2p_x - 2p_y}}\right]}{\cos^{-1}\left[\dfrac{1}{\sqrt{3}}\right]}$$

*All Evenness Measures for q=2*

Figure 3 illustrates that the angular distance-based evenness measure of order 2 ($^2E_7$) is more reactive to changes in relatively even assemblages than any other evenness measure within this framework.

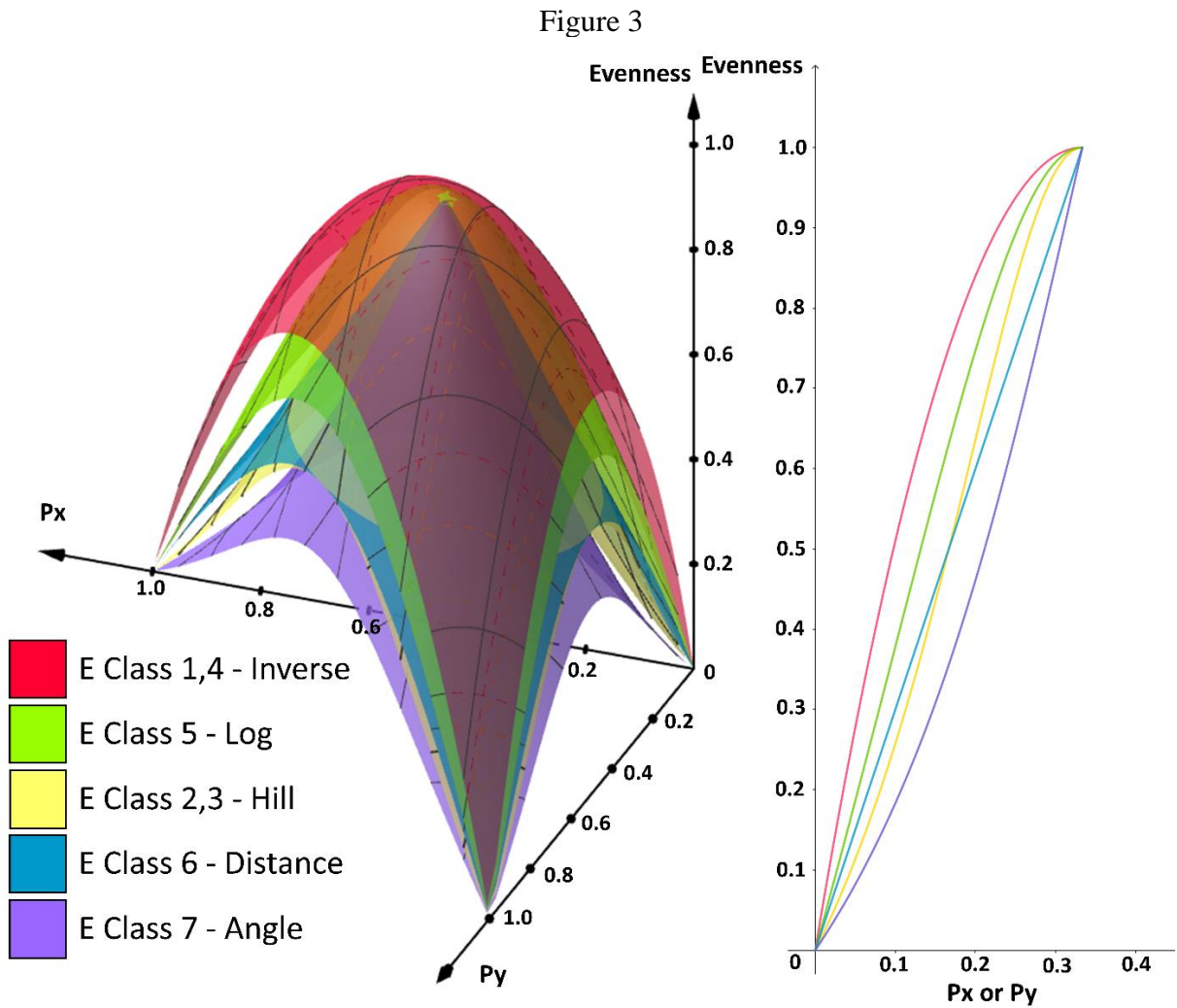

Figure 3

Figure 3. To the left each of the evenness measures of order 2 is plotted as a surface in three-dimensional space. To the right each of the evenness measures of order 2 is represented by a line, showing their values for the case where the x axis is equal to both the x and y components of the relative abundance vector.

Figure 3 also shows that the Euclidean distance-based measure, $^2E_6$, intersects with $^2E_{2,3}$ at the point where diversity equals (S-1), which means that the Euclidean distance does not fit neatly into our nesting of divergence-based measures. $^2E_6$ is more reactive than $^2E_{2,3}$ to changes in relatively even assemblages, but it is also more reactive than $^2E_{2,3}$ to changes in relatively uneven ones. It is less reactive near the middle of the range. Since the two measures intersect when diversity equals (S-1) larger values of S cause $^2E_6$ to cross below $^2E_{2,3}$ sooner, meaning that their relative reactivity depends on richness. Changes in richness also impact the relative reactiveness of the angular distance-based measure. Increasing richness causes $^2E_{2,3}$ to cross below $^2E_7$ in highly uneven assemblages, with this crossing point also moving closer to the mean vector as S increases.

So, while we can prove a ranking for the divergence-based measures, the relative reactivity of $^2E_6$ and $^2E_7$ compared to those measures depends on richness. We will illustrate this later using simulated ecological data, but for now it is important to note that larger values of richness complicate the process of directly ranking the distance-based evenness measures alongside the divergence-based ones.

*All Evenness Measures When q=1*

When q=1 it is no longer the case that measures of evenness are fully rotationally symmetric. Instead, the curve of equal diversity and evenness is the curve where the first Hill-number (the exponent of Shannon entropy) is a constant. That said, we can still visualize the evenness measures as surfaces when q=1. First, notice that choosing q=1 simplifies the five classes of divergence-based evenness measures to three:

$$^1E_{1,2,5} = \text{Pielou's J} = \frac{\ln(^1D)}{\ln(S)} = \frac{-\sum_{i=1}^{S} p_i \ln p_i}{\ln(S)}$$

$$^1E_3 = \text{Heip Evenness} = \frac{^1D - 1}{S - 1}$$

$$^1E_4 = \frac{1 - (1/\,^1D)}{1 - (1/S)}$$

The distance-based measure of order 1 is the Bulla (1994) measure which does not exhibit the principle of transfer (since its derivative is discontinuous), so the only remaining evenness measure is the angular distance-based measure:

$$^1E_7 = 1 - \frac{\cos^{-1}\left[\frac{^1D}{S}\right]}{\cos^{-1}\left[\frac{1}{S}\right]}$$

By setting S=3 and recognizing that valid relative abundance vectors must sum to one, diversity of order 1 can be written as:

$$^1D = e^{-[p_x \ln p_x + p_y \ln p_y + (1 - p_x - p_y)\ln(1 - p_x - p_y)]}$$

which we can plug back into the expression for each evenness measure to create four surfaces:

$$^1E_{1,2,5} = \frac{p_x \ln p_x + p_y \ln p_y + (1 - p_x - p_y)\ln(1 - p_x - p_y)}{\ln(3)}$$

$$^1E_3 = \frac{e^{-[p_x \ln p_x + p_y \ln p_y + (1 - p_x - p_y)\ln(1 - p_x - p_y)]} - 1}{2}$$

$$^1E_4 = \frac{1 - e^{[p_x \ln p_x + p_y \ln p_y + (1 - p_x - p_y)\ln(1 - p_x - p_y)]}}{2/3}$$

$$^1E_7 = 1 - \frac{\cos^{-1}\left[\frac{e^{-[p_x \ln p_x + p_y \ln p_y + (1 - p_x - p_y)\ln(1 - p_x - p_y)]}}{3}\right]}{\cos^{-1}\left[\frac{1}{3}\right]}$$

As Figure 4 shows, these four surfaces also nest within each other. The appendix proves that this relationship holds over all values of richness, and they can therefore be ranked by their reactivity to changes in relatively uneven assemblages as:

$$^1E_4 > {}^1E_{1,2,5} > {}^1E_3 > {}^1E_7$$

with the opposite ordering once again holding for relatively even assemblages.

Figure 4

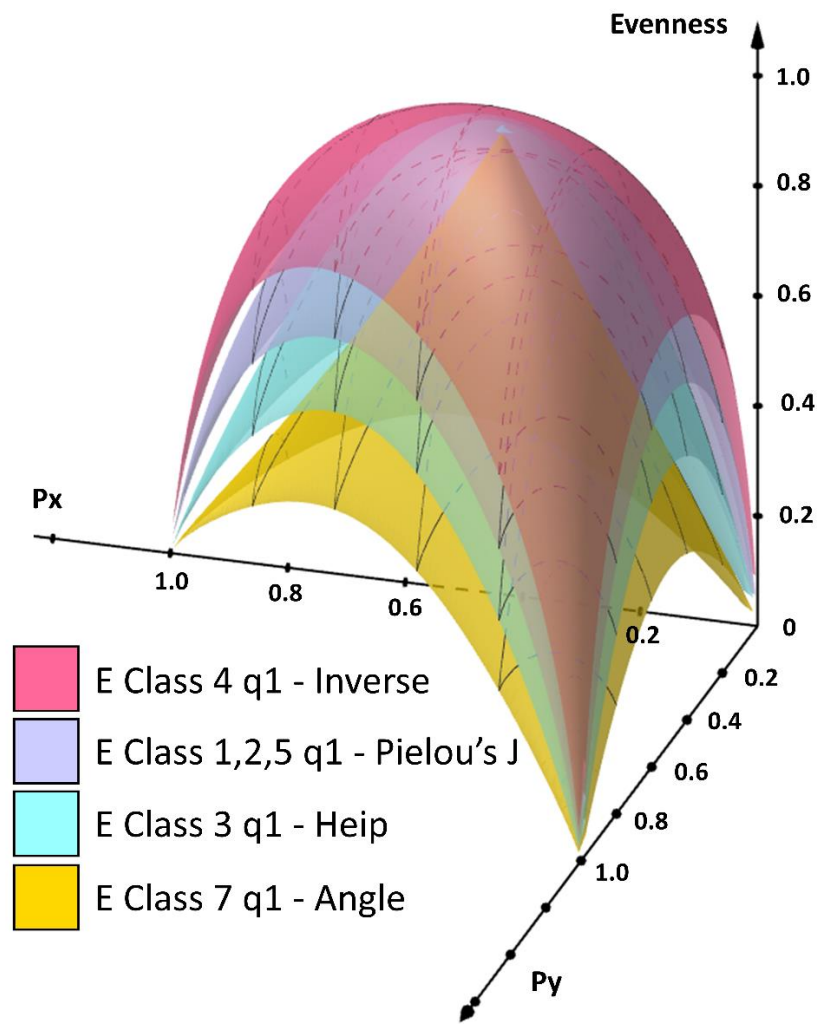

Figure 4. The four evenness measures of order 1 are plotted as surfaces to show that they also nest within each other when richness is held constant. The two-dimensional view is not illustrative in this case because the order 1 measures are not rotationally symmetric.

AN ILLUSTRATION OF THESE RELATIONSHIPS FOR SIMULATED DATA

Following Faith and Du (2018), we illustrate these relationships for simulated species abundance data created using the geometric distribution. While it is reasonable to argue that some other distribution might be a closer fit to real species abundance data, as Faith and Du (2018) argue, the geometric distribution is a reasonable choice given its natural skew and has the advantage of using only a single parameter to control its shape.

The geometric distribution gives the abundance of the $i$-th species from a parameter $k$, which must be larger than 0 and smaller than 1, as:

$$a_i = k \cdot (1-k)^{i-1}$$

For any reasonably large $k$ or S, the sum of these abundances will be 1, and so they can be treated as proportional abundances. When both $k$ and S are small however the resulting abundances must be scaled by their sum. We can smoothly vary the evenness of our simulated assemblages using this distribution since small values of k create highly even assemblages and large values of k create highly uneven ones. Then, by varying number of species we include in our calculations we can simultaneously manipulate the value of S, allowing us to observe how the relationships between these evenness measures change with richness. Table 1 shows a summary of these results.

Table 1

| Class | Name | S = 5<br>k = 0.1 | More Even<br>0.2 | 0.3 | 0.4 | 0.5 | 0.6 | 0.7 | 0.8 | Less Even<br>0.9 |
|---|---|---|---|---|---|---|---|---|---|---|
| $^2D$ | q=2 Diversity | 4.89 | 4.56 | 4.04 | 3.42 | 2.82 | 2.29 | 1.85 | 1.50 | 1.22 |
| $^2E_{1,4}$ | q=2 Inverse | 99.45% | 97.57% | 94.03% | 88.48% | 80.65% | 70.32% | 57.36% | 41.61% | 22.73% |
| $^2E_5$ | q=2 Log | 98.64% | 94.24% | 86.69% | 76.45% | 64.38% | 51.37% | 38.16% | 25.15% | 12.47% |
| $^2E_{2,3}$ | q=2 Hill | 97.30% | 88.94% | 75.90% | 60.57% | 45.45% | 32.15% | 21.20% | 12.48% | 5.55% |
| $^2E_6$ | q=2 Distance | 92.57% | 84.42% | 75.56% | 66.06% | 56.01% | 45.52% | 34.70% | 23.59% | 12.09% |
| $^2E_7$ | q=2 Angle | 86.68% | 72.72% | 58.94% | 46.13% | 34.82% | 25.19% | 17.15% | 10.46% | 4.83% |

| Class | Name | k = 0.1 | 0.2 | 0.3 | 0.4 | 0.5 | 0.6 | 0.7 | 0.8 | 0.9 |
|---|---|---|---|---|---|---|---|---|---|---|
| $^1D$ | q=1 Diversity | 4.95 | 4.76 | 4.45 | 4.00 | 3.47 | 2.90 | 2.35 | 1.86 | 1.43 |
| $^1E_4$ | q=1 Inverse | 99.72% | 98.76% | 96.88% | 93.75% | 88.93% | 81.87% | 71.87% | 57.93% | 37.88% |
| $^1E_{1,2,5}$ | q=1 Pielou's J | 99.32% | 97.00% | 92.70% | 86.13% | 77.22% | 66.11% | 53.16% | 38.68% | 22.44% |
| $^1E_3$ | q=1 Heip | 98.63% | 94.11% | 86.14% | 74.99% | 61.63% | 47.45% | 33.82% | 21.59% | 10.87% |
| $^1E_7$ | q=1 Angle | 89.18% | 77.50% | 65.29% | 53.00% | 41.21% | 30.44% | 21.07% | 13.19% | 6.55% |
| | S = 15 | More Even | | | | | | | | Less Even |
| Class | Name | k = 0.1 | 0.2 | 0.3 | 0.4 | 0.5 | 0.6 | 0.7 | 0.8 | 0.9 |
| $^2D$ | q=2 Diversity | 12.51 | 8.39 | 5.61 | 4.00 | 3.00 | 2.33 | 1.86 | 1.50 | 1.22 |
| $^2E_{1,4}$ | q=2 Inverse | 98.58% | 94.37% | 88.05% | 80.33% | 71.43% | 61.22% | 49.45% | 35.71% | 19.48% |
| $^2E_5$ | q=2 Log | 93.30% | 78.54% | 63.70% | 51.16% | 40.57% | 31.29% | 22.86% | 14.97% | 7.41% |
| $^2E_{2,3}$ | q=2 Hill | 82.23% | 52.77% | 32.95% | 21.40% | 14.28% | 9.52% | 6.12% | 3.57% | 1.59% |
| $^2E_6$ | q=2 Distance | 88.08% | 76.27% | 65.44% | 55.65% | 46.55% | 37.73% | 28.90% | 19.82% | 10.27% |
| $^2E_7$ | q=2 Angle | 67.97% | 44.56% | 30.32% | 21.47% | 15.46% | 11.02% | 7.51% | 4.63% | 2.16% |
| $^1D$ | q=1 Diversity | 13.61 | 10.42 | 7.43 | 5.36 | 4.00 | 3.07 | 2.39 | 1.87 | 1.44 |
| $^1E_4$ | q=1 Inverse | 99.27% | 96.86% | 92.73% | 87.14% | 80.35% | 72.24% | 62.37% | 49.82% | 32.48% |
| $^1E_{1,2,5}$ | q=1 Pielou's J | 96.40% | 86.56% | 74.07% | 61.98% | 51.18% | 41.42% | 32.22% | 23.10% | 13.34% |
| $^1E_3$ | q=1 Heip | 90.05% | 67.32% | 45.95% | 31.12% | 21.42% | 14.79% | 9.95% | 6.21% | 3.11% |
| $^1E_7$ | q=1 Angle | 71.11% | 46.65% | 30.03% | 19.85% | 13.50% | 9.27% | 6.22% | 3.87% | 1.93% |
| | S = 50 | More Even | | | | | | | | Less Even |
| Class | Name | k = 0.1 | 0.2 | 0.3 | 0.4 | 0.5 | 0.6 | 0.7 | 0.8 | 0.9 |
| $^2D$ | q=2 Diversity | 18.81 | 9.00 | 5.67 | 4.00 | 3.00 | 2.33 | 1.86 | 1.50 | 1.22 |
| $^2E_{1,4}$ | q=2 Inverse | 96.61% | 90.70% | 84.03% | 76.53% | 68.03% | 58.31% | 47.10% | 34.01% | 18.55% |
| $^2E_5$ | q=2 Log | 75.00% | 56.17% | 44.34% | 35.44% | 28.08% | 21.66% | 15.82% | 10.36% | 5.13% |
| $^2E_{2,3}$ | q=2 Hill | 36.34% | 16.33% | 9.52% | 6.12% | 4.08% | 2.72% | 1.75% | 1.02% | 0.45% |
| $^2E_6$ | q=2 Distance | 81.60% | 69.51% | 60.04% | 51.55% | 43.46% | 35.43% | 27.26% | 18.77% | 9.75% |
| $^2E_7$ | q=2 Angle | 36.27% | 20.73% | 14.10% | 10.14% | 7.39% | 5.31% | 3.64% | 2.25% | 1.06% |
| $^1D$ | q=1 Diversity | 24.99 | 12.20 | 7.66 | 5.38 | 4.00 | 3.07 | 2.39 | 1.87 | 1.44 |
| $^1E_4$ | q=1 Inverse | 97.96% | 93.68% | 88.72% | 83.07% | 76.53% | 68.80% | 59.40% | 47.45% | 30.93% |
| $^1E_{1,2,5}$ | q=1 Pielou's J | 82.27% | 63.95% | 52.05% | 43.01% | 35.44% | 28.67% | 22.31% | 15.99% | 9.23% |
| $^1E_3$ | q=1 Heip | 48.95% | 22.87% | 13.59% | 8.94% | 6.12% | 4.22% | 2.84% | 1.77% | 0.89% |
| $^1E_7$ | q=1 Angle | 32.45% | 14.61% | 8.63% | 5.66% | 3.87% | 2.67% | 1.80% | 1.12% | 0.56% |

Table 1. This table reports the calculated values of each diversity and evenness measure we consider for a wide range of evenness and richness. In every case these measures were calculated using simulated species abundance data following a geometric distribution where the parameter k varies inversely with diversity and evenness.

The first thing to note in table 1 is that the ordering of the divergence-based evenness measures we discussed above is clearly reflected in the simulated data. Regardless of the value of species richness used, the three divergence-based evenness measures of order q=2 ($^2E_{1,4}$, $^2E_5$, and $^2E_{2,3}$) cleanly nest. As we move from the left to the right in Table 1 evenness decreases, and $^2E_{1,4}$ always remains above $^2E_5$ which always remains above $^2E_{2,3}$. The numerical differences between these three measures are important and become more pronounced as S increases. When diversity $^2D$ is close to two, $^2E_{1,4}$ reports an evenness above 50% across the three richness levels tested. The other divergence-based measures report very different results. For example, when S=15, $^2E_5$ reports an evenness less than 30% and $^2E_{2,3}$ reports an evenness less than 10% for the same value of diversity. This illustrates how, even for quite normal circumstances, the divergence-based evenness measures can disagree markedly, with the least reactive measure reporting more than 50% evenness, while the most reactive measure reports less than 10% evenness.

The results for the order q=1 measures are quite similar. $^1E_4$ always takes a value above $^1E_{1,2,5}$, which always takes a value above $^1E_3$, which always takes a value above $^1E_7$. The differences between the measures are again quite pronounced and become amplified as S increases. For diversity $^1D$ close to 7.5 and S=50, for example, $^1E_4$ reports evenness of over 88%, while $^1E_{1,2,5}$ reports evenness near 50%, $^1E_3$ reports and evenness near 13%, and $^1E_7$ reports evenness of less than 9%. It seems difficult to accept that these measures all represent the same underlying concept, since some claim that evenness is very close to its maximum and others claim it is almost at its minimum for the same assemblage.

Table 1 also illustrates the fact that, when q=2, integrating the distance-based measures into this ranking is complex. For the low richness case where S=5 the pattern that we see is close to what we graphed earlier, with the angular-distance based measure $^2E_7$ remaining below every other measure and the distance-based measure $^2E_6$ crossing below the scaled Hill number $^2E_{2,3}$ near the middle of the range of possible evenness. For higher values of richness this crossing point moves further towards the case of perfect evenness however, and for the highest richness cases $^2E_6$ is effectively always below $^2E_{2,3}$. $^2E_7$ similarly becomes less reactive than $^2E_{2,3}$ as richness increases.

## A PATH FORWARD FOR MEASURING EVENNESS

In the context of those results, it is important to mention an additional finding that arose from our geometric analysis of evenness. Currently, any smooth, monotonically increasing function of diversity can use the general form provided by Chao and Ricotta (2019) to become a valid evenness. If a function increases with diversity, then it decreases with distance from the mean vector, satisfying the requirement of transfer (1a). If a function is smooth then it is continuous, and because diversity is itself symmetric, the function will also be permutation invariant (1b). Being bounded between 0 and 1 follows from the scaling in the formula given by Chao and Ricotta (2019) and establishes unrelatedness (3a). Using proportions results in scale invariance (3b). Rarity (2) requires a negative partial derivative with respect to S, which is harder to argue simply, but which we prove in the appendix also happens automatically when using Chao and Ricotta's (2019) suggested form. Because of this, under the current constraints, there are a theoretically infinite number of divergence-based evenness measures.

Thus, while Chao and Ricotta's (2019) unification of evenness is an important milestone for the field it cannot be the final word. We must do better than a theoretically infinite number of measures if we want avoid confusion and eventually settle on how best to operationalize the concept of evenness. There are likely two paths forward. Future work can either establish which measures to use empirically, or it can do so theoretically.

If we want to approach the question empirically then a simple suggestion might be to encourage studies to use and report multiple different evenness measures simultaneously. This effort might rightly receive pushback however, since it would encourage cherry-picking, and discourage researchers from understanding what aspect of diversity each study should be measuring (Magurran 2013 p.101). An alternative path that might mitigate this concern would be a large-scale replication effort. If a broad range of recent results pertaining to evenness could be reexamined using the full range of evenness measures, we might uncover patterns in which kind of divergence best represents the underlying ecology, potentially revealing unexpected nuances about how evenness functions in the world. The benefit of such an approach is that we would not need a breakthrough in our understanding of what evenness is, but the cost would be a significant effort spent on replicating research using many different evenness measures.

If we instead want to approach the question theoretically then our next step should be to narrow the allowable bounds for evenness as a function of diversity. This is in line with the sentiment expressed by Kvålseth (2015) in their discussion of value validity. A theory of evenness must establish an intuitive mapping between abundance distributions and the underlying concept of evenness. Kvålseth (2015) briefly argues that the most logical mapping is a Euclidean distance. Chao and Ricotta (2019) disagree. Kvålseth's (2015) solution is certainly elegant, but elegance is less persuasive than correctly reflecting something about the world. This is a central tension in the evenness literature, and carefully shedding light on which divergence best reflects the concept of evenness will be highly impactful.

In reference to $^2E_6$, and at several other times in their paper, Chao and Ricotta (2019) dismiss certain evenness measures for their tendency to overweight changes in dominant species, yet they stop short of creating any requirement for evenness based on this logic. Perhaps some traction could come from extending this thinking to assemblages rather than just species? Of course, we should not simply choose the measure which weights changes in even assemblages the least since an absurd measure that stayed as close as possible to 1 without violating the requirements of evenness would clearly not be reflective of the underlying concept. So there must be an upper bound on evenness as a function of diversity, but where is that upper bound?

A lower bound would be important as well and recognizing that $E_3$ is a normalized Hill-number might provide one. Consider this question: If the diversity of an assemblage reduces by 10%, does the evenness of that assemblage decrease by 10% also? If the answer is yes, then $E_3$ is the only correct reflection of evenness. If not, then the question becomes whether, in highly even (or uneven) assemblages, changes in diversity should change evenness by more than they change diversity, or by less. Answering that question will neatly restrict the range of valid divergences, and therefore the valid evenness measures. For example, if it were convincingly argued that evenness should be less reactive than diversity to changes in highly even assemblages, then $E_3$, the two distance-based families $E_6$ and $E_7$, and any future measures that are found to be more reactive than the Hill-numbers to changes in relatively even assemblages could all be excluded from consideration as reflections of evenness. The excluded measures could of course be re-cast as reflections of some yet unnamed aspect of diversity, but they would not be evenness. Agreeing on conceptual limits such as these would be contentious, but the benefits of a theoretical approach are the promise of a more precise understanding of evenness as a concept. Hopefully, the insights into evenness that we establish here will be the first steps on our path towards that understanding.

## DISCUSSION AND CONCLUSION

This geometric analysis of evenness provides a new perspective on the topic. On the one hand we reveal that the current requirements of an evenness measure leave room for a theoretically unlimited proliferation, but on the other hand we also show that, for a given richness S and order q=1 or q=2, the families of divergence-based evenness measures can be ranked based on their reactivity to changes in relatively even or relatively uneven assemblages. This lens can help us narrow our focus as we work towards a deeper understanding of evenness, whether we arrive at this understanding empirically or theoretically.

While this distinction may seem technical, we should emphasize that ranking families of measures based on their reactivity to changes in relatively even or relatively uneven assemblages is importantly different from ranking measures within a family based on their reactivity to dominant or rare species. Ecologists have long used order q to rank measures within each family based on their reactivity to dominant or rare species (Ricotta 2003). Indeed, this is the basis upon which Chao and Ricotta (2019) restrict their analysis to measures with order $q \leq 2$. Establishing relationships between families is novel however and represents one of the central contributions of this work.

These relationships can also be used to motivate empirical work and extract deeper insight from its results. Authors now have the option of justifying their choice of measure based on what theory tells them about where along the range of possible abundance distributions changes in evenness will be most important. If they believe that changes in the most even assemblages are most important then they should consider $E_7$, if they want to focus on changes in the most uneven ones then $E_4$ stands out. More balanced approaches could come from employing $E_6$ or $E_3$, with $E_6$ providing a measure that weights changes in evenness linearly, and $E_3$ weighting changes in evenness proportionally with changes in diversity. This significantly increases what we can observe, test, and understand about evenness. Of course, other considerations about the performance of these measures in practice, such as their sensitivity and stability in the face of incomplete sampling (Faith and Du 2018) are important. The fact that increasing richness widened the gap between the measures we considered here further reinforces that concern, because larger sample sizes mechanically increase observed richness.

In closing, nothing in this paper should be taken as an argument against the framework Chao and Ricotta (2019) provide. While we argue that there is still considerable work to be done before we understand what kind of divergence best represents evenness, we believe that the concept of a scaled divergence itself remains an excellent tool for operationalizing evenness. Establishing a relationship between the families of evenness measures in the way we present is simply an additional layer of understanding, and a lens that we hope will enable future research to view measures of evenness more clearly.

# Appendix

# Using Geometry to Rank Evenness Measures: Towards a Deeper Understanding of Divergence

*Appendix S1. Visualizing Measures of Evenness and Diversity Individually*

Figure S1 shows the diversity surface:

$$^2D = \frac{1}{2p_x^2 + 2p_y^2 + 2p_x p_y + 1 - 2p_x - 2p_y}$$

Figure S1

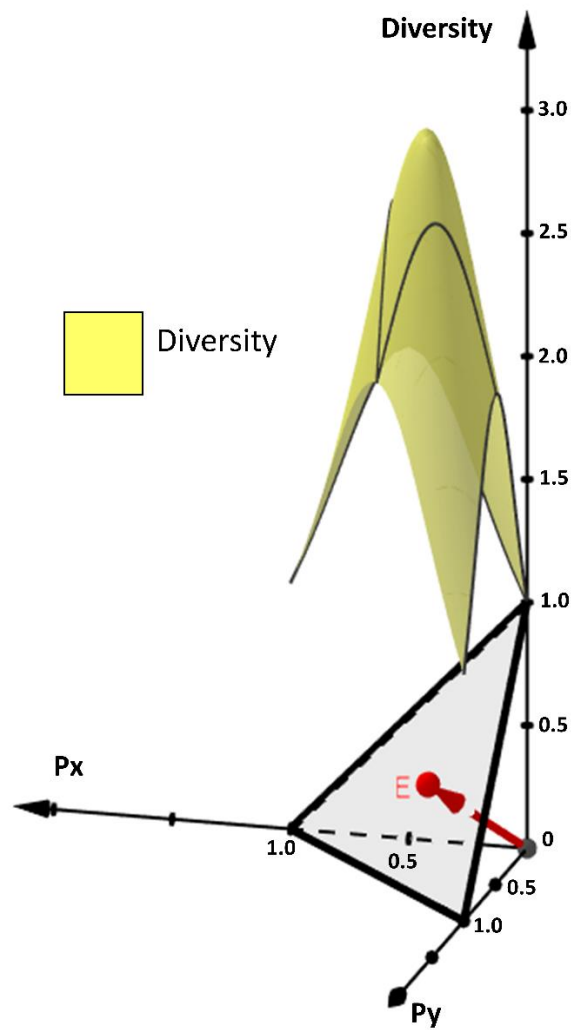

Figure S1. Diversity is function that maps each relative abundance vector onto a surface that ranges between 1 and S. This surface is show alongside the triangle of valid relative abundances, and the mean vector.

Though the triangle of vaid relative abnundances and the mean vector do not exist in the same space as the diversity or evenness surfaces, we include them in these visuals to help readers see how abundance vectors map to each surface. Figure S2 shows the surface:

$$^2E_{2,3} = \frac{1}{2} \cdot \left( \frac{1}{2p_x^2 + 2p_y^2 + 2p_x p_y + 1 - 2p_x - 2p_y} - 1 \right)$$

Figure S2

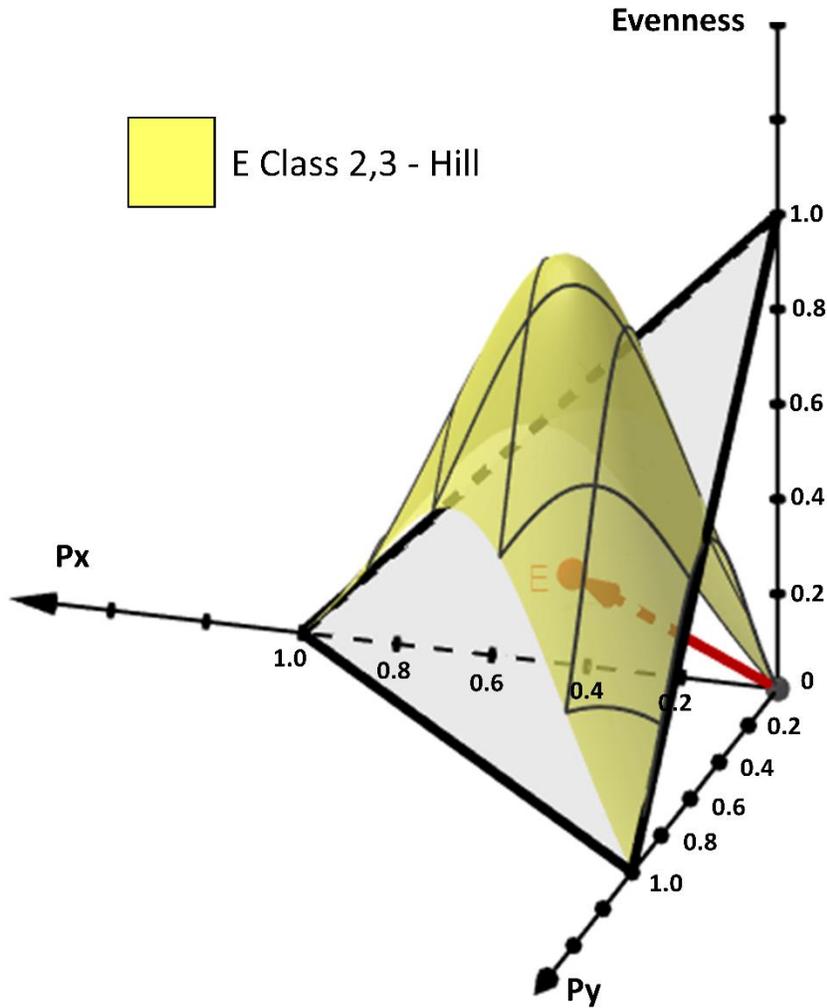

Figure S2. The class 2 and 3 evenness measures for q=2 shift diversity down and scale it to the range [0,1]. This surface shows the value of evenness returned by the class 2 and 3 evenness measures for every vector ending on the triangle of valid relative abundances.

When Chao and Ricotta (2019) say that their third class of evenness measures is "the normalized absolute slope of the Hill-number profile" they mean that it is literally a scaled copy of the diversity surface. Seeing this relationship graphically helps to emphasize rather than obfuscate the fact that $^2E_{2,3}$ is a very simple, linear transformation of a Hill-number. Chao and Ricotta have defined their requirement of independence with this in mind, arguing that a fixed range between 0 and 1 is enough for a measure to be independent of diversity.

Figure S3 shows the surface:

$$^2E_{1,4} = \frac{3}{2} \cdot [1 - (2p_x^2 + 2p_y^2 + 2p_xp_y + 1 - 2p_x - 2p_y)]$$

Figure S3

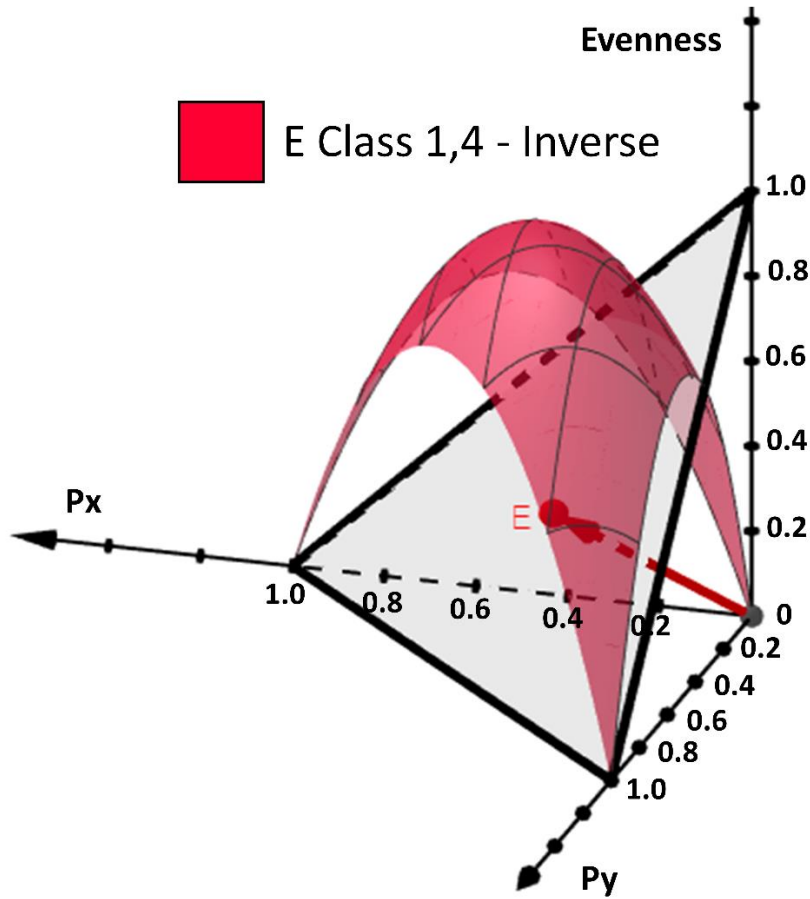

Figure S3. The class 1 and 4 evenness measure are based on the additive inverse of diversity. This surface shows the value of evenness returned by the class 1 and class 4 evenness measures for every vector ending on the triangle of valid relative abundances.

Figure S4 shows the surface:

$$^2E_5 = \frac{1}{\ln(3)} \cdot \ln\left(\frac{1}{2p_x^2 + 2p_y^2 + 2p_xp_y + 1 - 2p_x - 2p_y}\right)$$

Figure S4

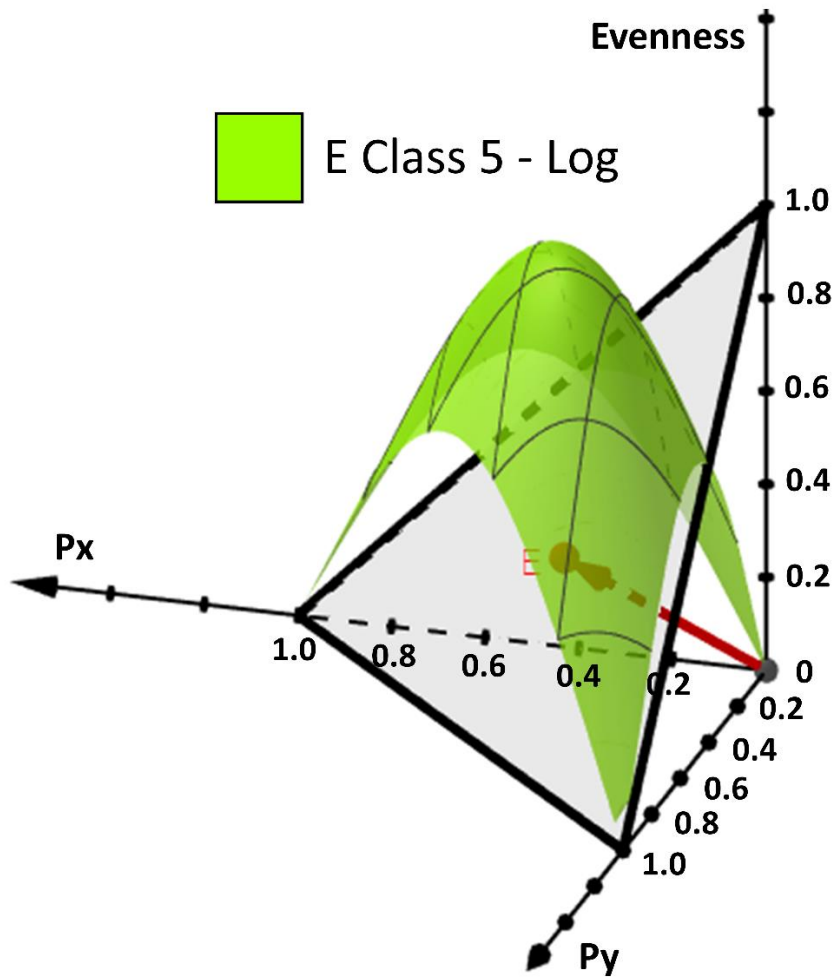

Figure S4. The class 5 evenness measure is a logarithmic transformation of diversity. This surface shows the value of evenness returned by the class 5 evenness measures for every vector ending on the triangle of valid relative abundances.

Figure S5 shows the surface:

$$^2E_6 = 1 - \frac{\sqrt{\left(p_x - \frac{1}{3}\right)^2 + \left(p_y - \frac{1}{3}\right)^2 + \left(1 - p_x - p_y - \frac{1}{3}\right)^2}}{\sqrt{\frac{2}{3}}}$$

Figure S5

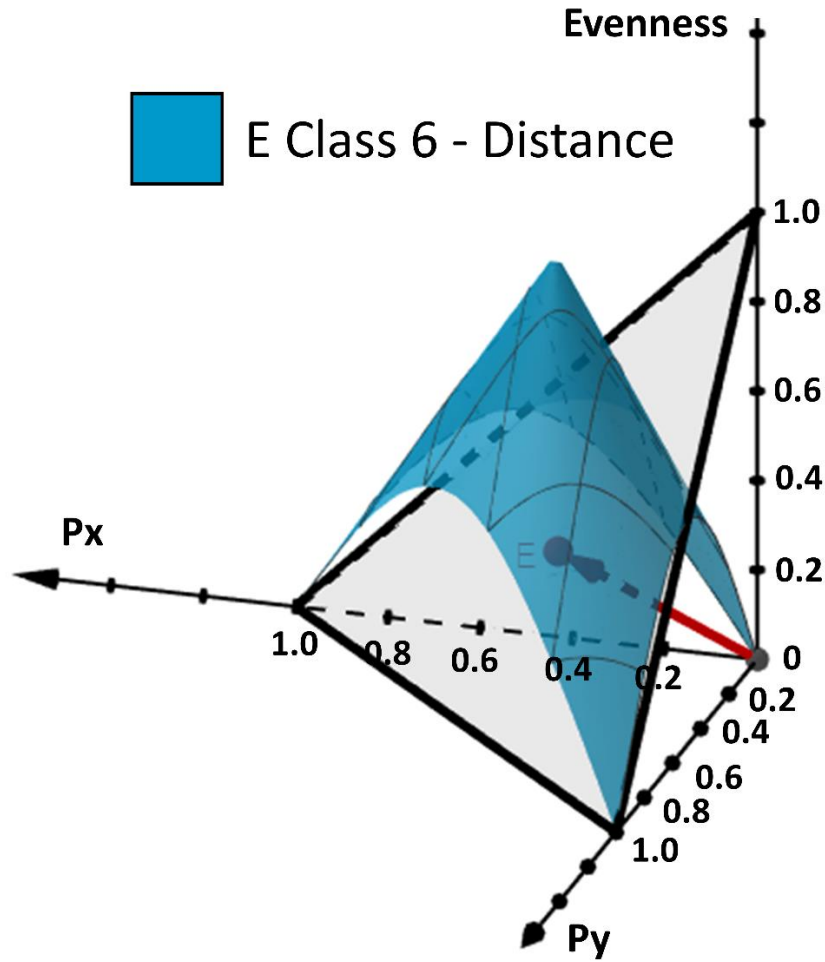

Figure S5. The distance-based class 6 evenness measure has a linear relationship with distance. This surface shows the value of evenness returned by the class 6 evenness measure for every vector ending on the triangle of valid relative abundances.

*Appendix S2. Deriving the Angular Distance Based Measure of Evenness*

We can measure the angle θ in figure 1 using the vectors **A** and **E**, since:

$$\cos\theta = \frac{\vec{A}\cdot\vec{E}}{\|\vec{A}\|\|\vec{E}\|} = \frac{\sum_{i=1}^{S}\frac{p_i}{S}}{\sqrt{\sum_{i=1}^{S}(p_i)^2}\cdot\sqrt{\sum_{i=1}^{S}\left(\frac{1}{S}\right)^2}} = \frac{\frac{1}{S}}{\sqrt{\frac{1}{S}\cdot\sum_{i=1}^{S}(p_i)^2}} = \frac{1}{\sqrt{S\cdot\sum_{i=1}^{S}(p_i)^2}}$$

and therefore:

$$\theta = \cos^{-1}\left[\left(\frac{1}{S\cdot\sum_{i=1}^{S}(p_i)^2}\right)^{1/2}\right] = \cos^{-1}\left[\left(\frac{1}{S}\cdot\frac{1}{\sum_{i=1}^{S}(p_i)^2}\right)^{1/2}\right] = \cos^{-1}\left[\left(\,^2D/S\right)^{1/2}\right]$$

The maximum possible angle between these vectors is formed when **p** approaches (1,0,…,0), and $^2D$ equals 1, so:

$$\theta_{max} = \cos^{-1}\left[(1/S)^{1/2}\right]$$

In the framework of Chao and Ricotta (2019) these two expressions are a divergence for q=2 and a scaling factor, enabling us to use the general form of an evenness measure to express the normalized angular distance as an evenness of order 2, which we label as the seventh class of evenness measure:

$$^2E_7 = 1 - \frac{\cos^{-1}\left[\left(\,^2D/S\right)^{1/2}\right]}{\cos^{-1}\left[(1/S)^{1/2}\right]}$$

We can expand this formulation to a general divergence of order q ≥ 2, though these measures are no longer strictly angular distances:

$$^q\Delta_7 = \cos^{-1}\left[\frac{\left(\,^qD\right)^{1/q}}{(S)^{1/q}}\right]$$

and an angular evenness measure of general order q:

$$^qE_7 = 1 - \frac{\cos^{-1}\left[\frac{(^qD)^{1/q}}{(S)^{1/q}}\right]}{\cos^{-1}\left[\frac{1}{(S)^{1/q}}\right]}$$

When q=1 this naturally becomes:

$$^1E_7 = 1 - \frac{\cos^{-1}\left[\frac{^1D}{S}\right]}{\cos^{-1}\left[\frac{1}{S}\right]}$$

This is a novel family of evenness measures in the ecology literature which provide an interesting alternative definition for distance.

Returning to the case where q=2 and S=3, our angular distance-based evenness measure also yields a surface:

$$^2E_7 = 1 - \frac{\cos^{-1}\left[\frac{1}{\sqrt{3}\cdot\sqrt{2p_x^2 + 2p_y^2 + 2p_xp_y + 1 - 2p_x - 2p_y}}\right]}{\cos^{-1}\left[\frac{1}{\sqrt{3}}\right]}$$

Figure S6 shows this surface.

Visual inspection of Figure S6 is mostly sufficient to show that the measure satisfies the properties of evenness discussed by Chao and Ricotta (2019). Any movement away from the maximum diversity reduces evenness, so the principle of transfer holds (1a). Since the surface is smooth and rotationally symmetric, it is continuous and permutation invariant (1b). Since it is bounded between 0 and 1 it is unrelated to diversity (3a) and since it uses proportions it is scale invariant (3b). The final requirement asks the measure to not increase under the addition of a vanishingly rare species, which we cannot show graphically, but which can be shown mathematically by taking the partial derivative of the evenness measure with respect to S.

$$\frac{\partial {}^2E_7}{\partial S} = \frac{\partial}{\partial S}\left\{1 - \frac{\cos^{-1}\left[\left({}^2D/S\right)^{1/2}\right]}{\cos^{-1}[(1/S)^{1/2}]}\right\} = -\frac{\partial}{\partial S}\left\{\frac{\cos^{-1}\left[\left({}^2D/S\right)^{1/2}\right]}{\cos^{-1}[(1/S)^{1/2}]}\right\}$$

Figure S6

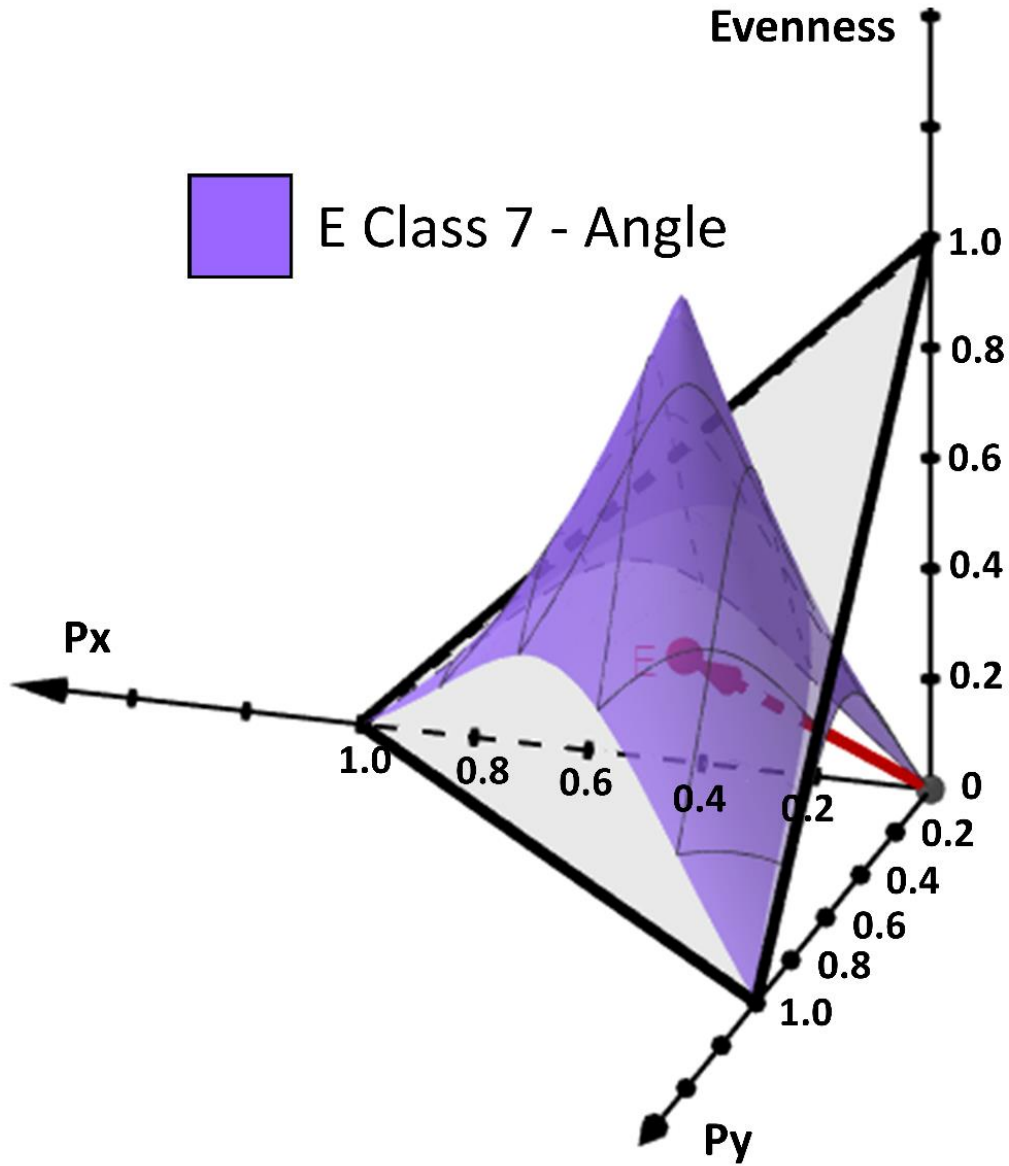

Figure S6. The angular distance-based class 7 evenness measure is more reactive to changes in dominant species than any other measure. This surface shows the value of evenness returned y the class 7 evenness measure for every vector ending on the triangle of valid relative abundances.

by the quotient rule:

$$-\frac{\partial}{\partial S}\left\{\frac{\cos^{-1}\left[(^2D/S)^{\frac{1}{2}}\right]}{\cos^{-1}\left[(1/S)^{\frac{1}{2}}\right]}\right\}$$

$$= -\frac{\frac{\partial}{\partial S}\left\{\cos^{-1}\left[(^2D/S)^{\frac{1}{2}}\right]\right\}\cdot \cos^{-1}\left[(1/S)^{\frac{1}{2}}\right] - \cos^{-1}\left[(^2D/S)^{\frac{1}{2}}\right]\cdot\frac{\partial}{\partial S}\left\{\cos^{-1}\left[(1/S)^{\frac{1}{2}}\right]\right\}}{\{\cos^{-1}[(1/S)^{1/2}]\}^2}$$

Since for any constant c:

$$\frac{\partial}{\partial S}\cos^{-1}\left[(c/S)^{\frac{1}{2}}\right] = \frac{-1}{\sqrt{1-\frac{c}{S}}}\cdot\frac{\partial}{\partial S}\left[\left(\frac{c}{S}\right)^{\frac{1}{2}}\right] = \frac{-1}{\sqrt{1-\frac{c}{S}}}\cdot\left(\frac{-c^{1/2}}{2S^{3/2}}\right) = \frac{c^{1/2}}{2S(S-c)^{1/2}}$$

we can work out that:

$$-\frac{\partial}{\partial S}\left\{\frac{\cos^{-1}\left[(^2D/S)^{\frac{1}{2}}\right]}{\cos^{-1}\left[(1/S)^{\frac{1}{2}}\right]}\right\}$$

$$= -\frac{\frac{^2D^{1/2}}{2S(S-\,^2D)^{1/2}}\cdot\cos^{-1}\left[(1/S)^{\frac{1}{2}}\right] - \frac{1}{2S(S-1)^{1/2}}\cdot\cos^{-1}\left[(^2D/S)^{\frac{1}{2}}\right]}{\{\cos^{-1}[(1/S)^{1/2}]\}^2}$$

Since $S > {}^2D > 1$, expect for the perfectly even case, and since $\cos^{-1}$ decreases over (0,1):

$$\cos^{-1}\left[(1/S)^{\frac{1}{2}}\right] > \cos^{-1}\left[(^2D/S)^{\frac{1}{2}}\right]$$

which means that:

$$\frac{^2D^{1/2}}{2S(S-\,^2D)^{1/2}}\cdot\cos^{-1}\left[(1/S)^{\frac{1}{2}}\right] > \frac{1}{2S(S-1)^{1/2}}\cdot\cos^{-1}\left[(^2D/S)^{\frac{1}{2}}\right]$$

So, the numerator in the calculation of the derivative is always positive, and the entire derivative is always negative, except for the perfectly even case when it is zero, which proves the requirement of rarity (2).

*Appendix S3. Proof of Nesting of Order 2 Divergence-Based Evenness Measures for S >3*

Proposition 1: $^2E_{1,4} > {}^2E_5 > {}^2E_{2,3}$ for all S > 3, and for all $^2D \in (1,S)$.

Since:

$$^2E_{1,4} = \frac{1 - (1/\,^2D)}{1 - (1/S)}$$

and

$$^2E_5 = \frac{\ln(\,^2D)}{\ln S}$$

if $^2E_{1,4} > {}^2E_5$ then:

$$\frac{1 - (1/\,^2D)}{1 - (1/S)} > \frac{\ln(\,^2D)}{\ln S}$$

grouping variables together this inequality can be written:

$$\frac{\ln S}{1 - (1/S)} > \frac{\ln(\,^2D)}{1 - (1/\,^2D)}$$

Next notice that $S > {}^2D$, since the range for diversity is between 1 and S. Now, because the function:

$$f(x) = \frac{\ln x}{1 - (1/x)}$$

has a strictly positive derivative over the positive real numbers, it monotonically increases with x, and since S is always greater than $^2D$ it follows that:

$$\frac{\ln S}{1 - (1/S)} > \frac{\ln(\,^2D)}{1 - (1/\,^2D)}$$

thus $^2E_{1,4} > {}^2E_5$ for all S > 3, and for all $^2D \in (1,S)$.

Now, since:

$$^2E_5 = \frac{\ln(\,^2D)}{\ln S}$$

and

$$^2E_{2,3} = \frac{^2D - 1}{S - 1}$$

if $^2E_5 > {}^2E_{2,3}$ then:

$$\frac{\ln(^2D)}{\ln S} > \frac{^2D - 1}{S - 1}$$

again, grouping variables algebraically, this will be true if:

$$\frac{\ln(^2D)}{^2D - 1} > \frac{\ln S}{S - 1}$$

Remember that $S > {}^2D$. Because the function:

$$g(x) = \frac{\ln x}{x - 1}$$

has a strictly negative derivative over the positive real numbers it is monotonically decreasing, and it follows that:

$$\frac{\ln(^2D)}{^2D - 1} > \frac{\ln S}{S - 1}$$

Therefore $^2E_5 > {}^2E_{2,3}$ for all $S > 3$, and for all $^2D \in (1,S)$.

Since $^2E_{1,4} > {}^2E_5$ for all $S > 3$, and for all $^2D \in (1,S)$, and since $^2E_5 > {}^2E_{2,3}$ for all $S > 3$, and for all $^2D \in (1,S)$, these expressions can be combined to say that $^2E_{1,4} > {}^2E_5 > {}^2E_{2,3}$ for all $S > 3$ and for all $^2D \in (1,S)$.

Q.E.D.

*Appendix S4. Proof of Nesting of Order 1 Evenness Measures for S >3*

For the most part, we can follow a similar logic as appendix S3 to establish the nesting of the order 1 evenness measures.

Proposition 2: $^1E_4 > {}^1E_{1,2,5} > {}^1E_3 > {}^1E_7$ for all $S > 3$, and for all $^1D \in (1,S)$.

Since:

$$^1E_4 = \frac{1 - (1/{}^1D)}{1 - (1/S)}$$

and

$$^1E_{1,2,5} = \frac{\ln({}^1D)}{\ln(S)}$$

the relative ordering for these expressions is clearly the same as in the order 2 case. That is:

$$\frac{\ln S}{1 - (1/S)} > \frac{\ln({}^1D)}{1 - (1/{}^1D)}$$

will hold because the function:

$$f(x) = \frac{\ln x}{1 - (1/x)}$$

has a strictly positive derivative over the positive real numbers, and $S > {}^1D$. Thus $^1E_4 > {}^1E_{1,2,5}$ for all $^1D \in (1,S)$.

Since:

$$^1E_3 = \frac{{}^1D - 1}{S - 1}$$

the next expression is once again identical to the order 2 case. We can establish that:

$$\frac{\ln({}^1D)}{{}^1D - 1} > \frac{\ln S}{S - 1}$$

because:

$$g(x) = \frac{\ln x}{x - 1}$$

has a strictly negative derivative over the positive real numbers and $S > {}^1D$. Thus ${}^1E_{1,2,5} > {}^1E_3$ for all ${}^1D \in (1,S)$.

The final required relationship is novel, however. Since:

$$^1E_7 = 1 - \frac{\cos^{-1}\left[\frac{^1D}{S}\right]}{\cos^{-1}\left[\frac{1}{S}\right]}$$

establishing that ${}^1E_3 > {}^1E_7$ is equivalent to establishing that:

$$\frac{^1D - 1}{S - 1} > 1 - \frac{\cos^{-1}\left[\frac{^1D}{S}\right]}{\cos^{-1}\left[\frac{1}{S}\right]}$$

or:

$$\frac{^1D - 1}{S - 1} + \frac{\cos^{-1}\left[\frac{^1D}{S}\right]}{\cos^{-1}\left[\frac{1}{S}\right]} > 1$$

To see that this is the case, first consider that the expression:

$$\frac{^1D - 1}{S - 1}$$

is bounded between 0 and 1 over the range of ${}^1D$, (1,S). When ${}^1D$ approaches 1 the value of the expression approaches 0 and when ${}^1D$ approaches S the value of the expression approaches 1. The rate of change of the expression with respect to diversity is equal to 1.

Now notice that the expression:

$$\frac{\cos^{-1}\left[\frac{^1D}{S}\right]}{\cos^{-1}\left[\frac{1}{S}\right]}$$

is also bounded between 0 and 1 over this range, but in a complementary way. Its value approaches 1 when $^1D$ approaches 1, and its value approaches 0 when $^1D$ approaches S.

The partial derivative of this expression with respect to diversity is

$$\frac{\partial}{\partial D}\left(\frac{\cos^{-1}\left[\frac{^1D}{S}\right]}{\cos^{-1}\left[\frac{1}{S}\right]}\right) = \frac{-1}{\cos^{-1}\left[\frac{1}{S}\right] \cdot \sqrt{S^2 - {^1D^2}}}$$

which, because S is larger than $^1D$, is negative and very close to zero until $^1D$ gets close enough to S that the denominator becomes less than 1. This only happens very close to perfect evenness.

Returning to the inequality in question, which is replicated below, when $^1D$ is at its minimum the first half of the left-hand side starts equal to 0 while the second half starts equal to 1. As $^1D$ grows, the first half increases in value at a rate of 1, and the second half decreases in value at a rate very close to 0. Thus, it must be the case that over the range over the range $^1D \in (1,S)$:

$$\frac{^1D - 1}{S - 1} + \frac{\cos^{-1}\left[\frac{^1D}{S}\right]}{\cos^{-1}\left[\frac{1}{S}\right]} > 1$$

and by extension, $^1E_3 > {^1E_7}$ over that same range. We can then combine these inequalities to state that $^1E_4 > {^1E_{1,2,5}} > {^1E_3} > {^1E_7}$ for all $S > 3$, and for all $^1D \in (1,S)$.

Q.E.D.

*Appendix S5. Proof of Rarity Following from the General Form of Evenness*

The property that Chao and Ricotta (2019) call rarity requires that "When a vanishingly rare species is added to an assemblage so that the diversity of order q > 0 barely changes, evenness should not increase." This is the same as saying that the partial derivative of evenness, with respect to richness, S, is negative. Using the general form of an evenness measure from their paper, we can write its partial derivative (or partial forward difference, since S is discrete) as:

$$\frac{\partial E}{\partial S} = \frac{\partial}{\partial S}\left[1 - \frac{\Delta(\boldsymbol{p},\overline{\boldsymbol{p}})}{\Delta(\boldsymbol{p}^0,\overline{\boldsymbol{p}})}\right] = -\frac{\partial}{\partial S}\left[\frac{\Delta(\boldsymbol{p},\overline{\boldsymbol{p}})}{\Delta(\boldsymbol{p}^0,\overline{\boldsymbol{p}})}\right]$$

where $\Delta$ is a valid divergence function, and $\overline{\boldsymbol{p}}$ is the vector of maximum evenness. Using the quotient rule to evaluate this derivative gives:

$$-\frac{\partial}{\partial S}\left[\frac{\Delta(\boldsymbol{p},\overline{\boldsymbol{p}})}{\Delta(\boldsymbol{p}^0,\overline{\boldsymbol{p}})}\right] = -\left\{\frac{\Delta(\boldsymbol{p}^0,\overline{\boldsymbol{p}}) \cdot \frac{\partial}{\partial S}[\Delta(\boldsymbol{p},\overline{\boldsymbol{p}})] - \Delta(\boldsymbol{p},\overline{\boldsymbol{p}}) \cdot \frac{\partial}{\partial S}[\Delta(\boldsymbol{p}^0,\overline{\boldsymbol{p}})]}{\Delta(\boldsymbol{p}^0,\overline{\boldsymbol{p}})^2}\right\}$$

Equation 3a from Chao and Ricotta (2019) defines a divergence for a function h as:

$$\Delta(\boldsymbol{p},\overline{\boldsymbol{p}}) = h(p_1^q, p_2^q, \ldots, p_S^q) - h[(1/S)^q, (1/S)^q, \ldots, (1/S)^q]$$

A rarity transformation assumes that diversity of order q barely changes with the addition of the rare species. This means that the $p_s$ added to the vector **p** during the rarity transformation is vanishingly small, and that the change in the first term of the definition above, $h(p_1^q, p_2^q, \ldots, p_S^q)$, is also vanishingly small. This logic allows us to use the definition of the forward difference operator to say that:

$$\frac{\partial}{\partial S}[\Delta(\boldsymbol{p},\overline{\boldsymbol{p}})] \approx h(p_1^q, p_2^q, \ldots, p_S^q) - h\{[1/(S+1)]^q, [1/(S+1)]^q, \ldots, [1/(S+1)]^q\}$$

$$- h(p_1^q, p_2^q, \ldots, p_S^q) + h[(1/S)^q, (1/S)^q, \ldots, (1/S)^q]$$

$$= -h\{[1/(S+1)]^q, [1/(S+1)]^q, \ldots, [1/(S+1)]^q\}$$

$$+ h\{[1/(S)]^q, [1/(S)]^q, \ldots, [1/(S)]^q\}$$

Similarly, $\mathbf{p}^0$ is defined as the vector where all elements, except for the first, are vanishingly small, and so by a similar line of argument we can say that:

$$\frac{\partial}{\partial S}[\Delta(\mathbf{p}^0,\overline{\mathbf{p}})] \approx -h\{[1/(S+1)]^q, [1/(S+1)]^q, \ldots, [1/(S+1)]^q\}$$

$$+ h\{[1/(S)]^q, [1/(S)]^q, \ldots, [1/(S)]^q\}$$

Since these two expressions are identical, it must be the case that, for a rarity transformation:

$$\frac{\partial}{\partial S}[\Delta(\mathbf{p},\overline{\mathbf{p}})] \approx \frac{\partial}{\partial S}[\Delta(\mathbf{p}^0,\overline{\mathbf{p}})]$$

Returning to the partial derivative:

$$-\frac{\partial}{\partial S}\left[\frac{\Delta(\mathbf{p},\overline{\mathbf{p}})}{\Delta(\mathbf{p}^0,\overline{\mathbf{p}})}\right] = -\left\{\frac{\Delta(\mathbf{p}^0,\overline{\mathbf{p}}) \cdot \frac{\partial}{\partial S}[\Delta(\mathbf{p},\overline{\mathbf{p}})] - \Delta(\mathbf{p},\overline{\mathbf{p}}) \cdot \frac{\partial}{\partial S}[\Delta(\mathbf{p}^0,\overline{\mathbf{p}})]}{\Delta(\mathbf{p}^0,\overline{\mathbf{p}})^2}\right\}$$

notice that $\mathbf{p}^0$ is the vector of maximum divergence, so:

$$\Delta(\mathbf{p}^0,\overline{\mathbf{p}}) > \Delta(\mathbf{p},\overline{\mathbf{p}})$$

and by combining the previous two insights, it must be the case that:

$$\Delta(\mathbf{p}^0,\overline{\mathbf{p}}) \cdot \frac{\partial}{\partial S}[\Delta(\mathbf{p},\overline{\mathbf{p}})] > \Delta(\mathbf{p},\overline{\mathbf{p}}) \cdot \frac{\partial}{\partial S}[\Delta(\mathbf{p}^0,\overline{\mathbf{p}})]$$

Therefore, the partial derivative (or partial forward difference) of evenness, E, with respect to S, will be negative for any evenness constructed from a valid divergence and following the general form provided by Chao and Ricotta (2019). This proves that the property of rarity will automatically arise using their suggested framework.

Q.E.D.

As an aside, discussion above also indicates that, as a practical matter, adding a single individual of a rare species will not always result in a reduction of evenness. If the assemblage is formed from a sufficiently small number of individuals and $\mathbf{p}$ starts off as sufficiently uneven, then adding a single individual of a rare type will increase diversity and divergence by enough that evenness will increase. Chao and Ricotta (2019) define rarity as only applying for the idealized case where diversity barely changes, but in practice individuals are quantized and so users of evenness measures should not be alarmed if adding a single individual of a rare type sometimes actually increases evenness.